\documentclass[preprint,superscriptaddress,nofootinbib]{revtex4}
\usepackage[dvips]{graphicx}
\usepackage{amsmath,amssymb}

\def\lromn#1{\uppercase\expandafter{\romannumeral#1}}

\begin{document}

  \preprint{UT-HET-083}

  \date{\today}

  \title{Proton stability in low-scale extra-dimensional
  grand unified theories} 
  \author{Mitsuru Kakizaki}
  \email{kakizaki@sci.u-toyama.ac.jp}
  \affiliation{
  Department of Physics,
  University of Toyama, Toyama 930-8555, Japan
  }
  \begin{abstract}
    We propose an extra-dimensional mechanism that adequately
    suppresses the rate of proton decay mediated by the $X$-bosons in
    grand unified theories.  The $X$-bosons and their Kaluza-Klein
    modes are localized in the extra spatial dimensions due to the
    kink configuration of the adjoint Higgs multiplet that is
    responsible for the spontaneous breaking of the unified gauge
    group while the standard model gauge bosons freely propagate in
    the bulk.  By localizing fermion multiplets apart from the
    $X$-bosons in the extra dimensions, the resulting four-dimensional
    couplings that give rise to proton decay are exponentially
    suppressed, leading to the longevity of proton even when the mass
    scale of the $X$-bosons is much lower than the usual unification
    scale.
  \end{abstract}
\maketitle

The discovery of a particle whose properties are consistent with the
Standard Model (SM) Higgs boson at the CERN Large Hadron Collider
(LHC) has completed the SM as a low-energy effective theory
\cite{higgs}.  Although the SM has successfully explained almost all
phenomena, the complicated structure of the SM gauge bosons and
fermions is still unexplained.

Grand unified theories (GUTs) provide a unified description of
elementary particles and their interactions in the SM
and its extensions \cite{Georgi:1974sy}.  The observed pattern of the
three SM gauge coupling constants can be naturally explained by
assuming the existence of hypothetical grand unification at some high
energy scale \cite{Georgi:1974yf}.  The electric charge operator
composed of the generators of a larger (semi-)simple group leads to
the quantization of the electric charges of the SM particles, which
are embedded into irreducible representations of the grand unified group.
The idea of grand unification makes an exciting prediction that the
proton is not absolutely stable \cite{review}.  Proton decay is a
consequence of the quark-lepton transition induced by the exchange of
the GUT-scale gauge bosons (referred to as
the $X$-bosons), which are required in order to complete the gauge
multiplet of the grand unified group.  The predicted proton decay rate
is estimated to be
\begin{eqnarray}
  \Gamma_p \sim g_X^4 \frac{m_p^5}{M_X^4}\, ,
\end{eqnarray}
where $m_p$ and $M_X$ are the proton and $X$-boson masses,
respectively, and $g_X$ is the coupling constant of the $X$-boson
interaction with the SM fermions.  However, proton decay has not been
observed.  The lower bound on the partial lifetime for $p \to e^+ \pi$,
the main $X$-boson-mediated decay mode in conventional GUTs,
is set by the Super-Kamiokande experiment \cite{SK}:
\begin{eqnarray}
  \tau_p(p \to e^+ \pi) > 8.2 \times 10^{33}~{\rm years}\, ,
  \quad (90\%~ {\rm C.L.})\, .
\end{eqnarray}
This limit is converted to the lower bound on the GUT scale:
\begin{eqnarray} \label{eq:bound}
  M_X/g_X \gtrsim 10^{15}~{\rm GeV}\, .
\end{eqnarray}
Theoretically, the values for $M_X$ and $g_X$ are predicted by solving
renormalization group equations.  In non-supersymmetric GUTs, the
$X$-boson mass is typically smaller than $10^{15}~{\rm GeV}$ and $g_X
\sim {\cal O}(1)$, conflicting with the results of the proton decay
experiment.  In supersymmetric (SUSY) GUTs, the GUT scale is pushed up
to $10^{16}$ GeV, suppressing adequately the $X$-boson-mediated proton
decay rate.  However, the exchange of the color triplet partners of
the SM Higgs doublets induces new operators that mediate proton decay
through superparticle-loop diagrams \cite{dimension5}.  Generically,
the colored-Higgs-mediated proton decay rate is less suppressed and
imposes a strong constraint on SUSY GUTs \cite{SUSYGUT}.  Interesting
ideas about proton decay have been proposed not only in usual
four-dimensional models but also in extra-dimensional models.
Orbifold constructions of SUSY GUTs eliminate colored-Higgs-mediated
proton decay, and can enhance the $X$-boson-mediated proton decay rate
to an observable level, depending on the matter field location
\cite{orbifold,hebecker}.  An alternative approach is to localize
chiral multiplets in extra dimensions using the kink configuration of
topological defects
\cite{Jackiw:1975fn,Rubakov:1983bz,ArkaniHamed:1999dc}.  In such a
context, separation of the colored-Higgs multiplets from the matter
multiplets leads to exponentially small overlaps of the wave
functions, and therefore colored-Higgs-mediated proton decay can be
suppressed to a negligible level \cite{KY1,Maru}.

In this paper, we propose a new mechanism for suppressing
$X$-boson-mediated proton decay by localizing the $X$-bosons apart
from SM fermions in extra dimensions.  We assume that the vacuum
expectation value (vev) of an adjoint GUT Higgs multiplet $\Sigma$ has
a non-trivial configuration in the extra dimensions, as proposed in
Ref.\cite{KY1}.  Given this assumption, we find that the gauge bosons
corresponding to the broken generators obtain mass, and are localized
at the zero of $\langle \Sigma \rangle$, like the chiral fermion
localization at the zero of a domain wall field.  On the other hand,
the SM gauge bosons remain massless and freely propagate in the bulk.
In the resulting four-dimensional low energy effective theory, the GUT
gauge invariance is not manifest.  The couplings between the
$X$-bosons and SM fermions, which are determined by the overlaps of
their wave functions, can have completely different values from the SM
gauge couplings.  Due to the exponential profiles of the wave
functions, the expectation that the $X$-boson mass should be larger
than $10^{15}$ GeV is violated.  In this letter, we concentrate on an
$SU(5)$ construction in five-dimensional space-time for simplicity,
and exemplify how the gauge field localization and proton decay
suppression are achieved.  We will see that a GUT singlet domain wall
field is also introduced in order to produce the correct chiralities
for the SM fermions.

First we consider $SU(5)$ GUT symmetry breaking caused by the kink
configuration of an adjoint Higgs field
${\Sigma(24)^A}_B~(A,B=1,\cdots,5)$ in five-dimensional
space-time, whose coordinates are denoted by $x^M=(x^\mu,
x^5)~(\mu=0,1,2,3)$.  The non-trivial field configuration of $\Sigma$
is also responsible for violation of higher dimensional translation
invariance, which is a source for localization of fields.  Let us consider the
following $Z_2$ symmetric Lagrangian for $\Sigma$:
\begin{eqnarray}
  {\cal L}_5 = {\rm tr} ( \partial^M \Sigma)(\partial_M \Sigma)
  + \frac{M_\Sigma^2}{2} {\rm tr} \Sigma^2
  - \frac{a}{4} ({\rm tr} \Sigma^2)^2 - \frac{b}{2} {\rm tr} \Sigma^4\, ,
\end{eqnarray}
where $M_\Sigma^2(> 0), a$ and $b$ are five-dimensional parameters.
The $SU(5)$ invariance and higher dimensional translation invariance are
simultaneously broken by assuming that the expectation value
satisfies the boundary conditions of
\begin{eqnarray} \label{eq:boundary}
  \langle \Sigma \rangle = 
  \left\{ 
    \begin{array}{ll}
      V {\rm diag} (2,2,2-3,-3)\, , & x^5 \to \infty\, ,\\
      - V {\rm diag} (2,2,2-3,-3)\, , & x^5 \to - \infty\, ,
    \end{array}
  \right.
\end{eqnarray}
where $V = M_\Sigma / \sqrt{2(15a+7b)}$.  Notice that there is no
$SU(5)$ rotation that transforms the latter to the former.
Normalizing the hypercharge component $\Sigma_Y$ as $\Sigma = \Sigma_Y
Y + \cdots$, with $Y = {\rm diag}(2,2,2,-3,-3)/(2\sqrt{15})$, the
solution of the equation of motion for $\Sigma_Y$ with the boundary
conditions (\ref{eq:boundary}) has a domain-wall profile:
\begin{eqnarray}
  \langle \Sigma_Y(x^5) \rangle = \sigma \tanh (m x^5)\, , ~
  \sigma = 2 \sqrt{15} V\, , ~ m = \frac{M_\Sigma}{2}.
\end{eqnarray}
Here, we set the kink at the origin of the $x^5$ axis.

Next we show that the gauge bosons corresponding to the broken
generators obtain mass and are localized at the zero of $\langle
\Sigma_Y \rangle$.  Since the GUT gauge boson multiplet couples to the
adjoint Higgs multiplet $\Sigma$, the $X$-bosons acquire a
position-dependent mass after the GUT symmetry breaking,
\begin{eqnarray}
  M_X^2(x^5) = 25 g_5^2 V^2 \tanh^2 (m x^5)\, ,
\end{eqnarray}
where $g_5$ is the five-dimensional $SU(5)$ gauge coupling constant.
The $X$-boson kinetic and mass terms of the five-dimensional Lagrangian
are
\begin{eqnarray}
  {\cal L}_5 = X^{\dag\mu}
  [ \partial^\nu \partial_\nu - \partial_5^2 + M_X^2 (x^5) ] X_\mu\, .
\end{eqnarray}
In the four-dimensional viewpoint, there appear a tower of massive
$X$-boson states.  We consider the case where the cutoff scale
$\Lambda$ is smaller than $5g_5V$ so that freely propagating
Kaluza-Klein (KK) modes of the $X$-bosons are eliminated: all KK
$X$-bosons are localized and discretized.  Otherwise, the proton lifetime
would be unacceptably short.  The KK mode wave functions
$f_X^{(n)}(x^5)$ and mass eigenvalues $M_{X^{(n)}}^2$ are found by
solving the eigenvalue equations of
\begin{eqnarray}
  [- \partial_5^2 +  M_X^2 (x^5) ]f_X^{(n)} (x^5) = 
  M_{X^{(n)}}^2 f_X^{(n)} (x^5)\, .
\end{eqnarray}
Due to the kink configuration of $\langle \Sigma_Y \rangle$, the wave
functions have exponential profiles centered at $x^5=0$:
\begin{eqnarray}
  f_X^{(n)} (\xi) & \sim & (1 - \xi^2)^{(c-n)/2} 
  {_2} F_1 \left(-n, 2c-n+1; c-n+1; \frac{1-\xi}{2} \right)\, ,
\nonumber \\
  n & = & 0,1,2, \cdots < c\, ,
\end{eqnarray}
where
\begin{eqnarray}
  \xi = \tanh (m x^5)\, , ~
  c = \frac{1}{2} \left( \sqrt{1 +
    \frac{100g_5^2V^2}{m^2} } - 1 \right) > 0\, ,
\end{eqnarray}
and $_2 F_1$ is a hypergeometric function.  The corresponding mass
eigenvalues are
\begin{eqnarray}
  M_{X^{(n)}}^2 = [ (2n+1)c - n^2 ] m^2 \, .
\end{eqnarray}
In particular, the normalized zero mode wave function is
\begin{eqnarray}
  f_X^{(0)} (x^5) = \frac{m^{1/2}}{\pi^{1/4}}
  \sqrt{\frac{\Gamma(c+1/2)}{\Gamma(c)}} \frac{1}{\cosh^c (m x^5)}\, ,
\end{eqnarray}
with the mass being $M_{X^{(0)}}^2 = c m^2$.  The rough profiles of
the wave functions are easily derived if we use the approximation of
$\tanh (m x^5) \simeq m x^5$.  In this approximation, the eigenvalue
equations are the same as those of the one-dimensional harmonic
oscillator system, and the wave functions are expressed in terms of
the Hermite polynomials.  The zero mode wave function has a Gaussian
profile,
\begin{eqnarray}
  f_X^{(0)} (x^5) = \left( \frac{\mu^2}{\pi} \right)^{1/4}
  {\rm exp} \left( - \frac{\mu^2 (x^5)^2}{2} \right)\, ,
\end{eqnarray}
with
\begin{eqnarray}
  \mu^2 = 5 g_5 V m\, .
\end{eqnarray}
The mass of the zero mode is given by $M_X^{(0)}=\mu$.  For the SM
gauge bosons, on the other hand, no five-dimensional mass term is
generated.  Therefore, the zero modes of the SM gauge boson fields
have flat wave functions and remain massless, ensuring the SM gauge
invariance.

Let us tern to the fermion sector.  In the minimal $SU(5)$ GUT, the
quark doublet $q_{L}$, the charge conjugated right-handed up-type
quark $u_{R}^c$, the charge conjugated right-handed down-type quark
$d_{R}^c$, the left-handed lepton doublet $l_{L}$ and the charge
conjugated right-handed lepton singlet $e_{R}^c$ are embedded into an
anti-quintuplet $\Phi(5^*) = (d_{R}^c, l_{L})$ and a decuplet
$\Psi(10) = (q_{L}, u_{R}^c, e_{R}^c)$ for each generation.
Notice that in five-dimensional space-time, the minimal spinor
representation is $4$-dimensional.  One way to obtain a chiral
zero mode fermion from a five-dimensional Dirac fermion is to couple
the fermion bilinear to the kink background of a scalar field.  In our
model, the Yukawa coupling of ${\rm tr} (\langle \Sigma \rangle
\bar{\Phi} \Phi)$ does not suffice for producing the correct chiral
structure of the SM fermions.  When $\langle \Sigma \rangle$ gives a
positive Dirac mass to the colored triplet, the Dirac mass of the
doublet is negative.  As a result, the triplet and doublet zero modes
have different chiralities.

\begin{figure}[t!]
    \includegraphics[scale=0.7]{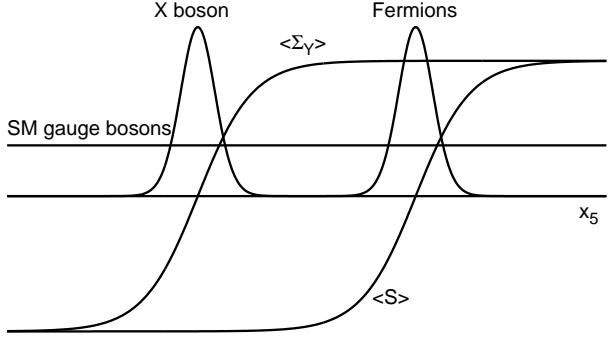}
    \caption{\label{fig:profile} \footnotesize Profiles of the
      zero mode wave functions of the $X$-bosons, SM gauge bosons and
      fermions.  The zero modes of the $X$-bosons (fermions) are
      localized at the zero of $\langle \Sigma_Y \rangle$ ($\langle S
      \rangle$).}
\end{figure}

To obtain the correct fermion chiralities, we
introduce a singlet scalar $S$ that has a domain wall profile,
\begin{eqnarray}
  \langle S (x^5) \rangle = v_S \tanh (m_S (x^5 - l) )\, ,
\end{eqnarray}
where $v_S$ is the vev of $S$ at $x^5 \to \infty$.  The
five-dimensional Lagrangian includes the Yukawa coupling of $\Phi(5^*)$ to the background fields $\Sigma$ and $S$:
\begin{eqnarray}
  {\cal L}_5 
  = {\rm tr} (\bar{\Phi}
    [i \gamma^\mu \partial_\mu - \gamma_5 \partial_5 - M_\Phi] \Phi)
  + \lambda_\Sigma {\rm tr}(\Sigma \bar{\Phi} \Phi)
    + \lambda_S S\ {\rm tr} (\bar{\Phi} \Phi) \, .
\end{eqnarray}
where $\gamma_5 = i \gamma^0 \gamma^1 \gamma^2 \gamma^3$.  Here,
$M_\Phi, \lambda_\Sigma$ and $\lambda_S$ are five-dimensional
parameters.  The Yukawa interaction terms for $\Psi(10)$ have
similar forms.  We assume that the domain wall profiles are stable
enough although their couplings to the fermion bilinears violate the
$Z_2$ invariance.  For simplicity, we choose $M_\Phi=0$ and omit
$\lambda_\Sigma$, which is a source of the $SU(5)$ violating profiles
of the quarks and leptons.  Then, all fermions are localized at the
zero of $\langle S \rangle$, and their wave functions have the same
profile.  The left-handed (right-handed) KK fermion wave functions
$f_L^{(n)}$ ($f_R^{(n)}$) are determined by the following coupled
equations:
\begin{eqnarray}
  \left[ - \partial_5 + M_F(x^5) \right] f^{(n)}_L(x^5)
  & = & m^{(n)} f_R^{(n)}(x^5)\, ,
  \nonumber \\
  \left[ \partial_5 + M_F(x^5) \right] f^{(n)}_R(x^5)
  & = & m^{(n)} f_L^{(n)}(x^5)\, ,
\end{eqnarray}
where $M_F(x^5) = - \lambda_S \langle S \rangle$.  For $\lambda_S >
0$, a left-handed chiral zero mode survives and the wave functions of
their KK fermions have the same form as those of the $X$-bosons,
\begin{eqnarray}
  f_L^{(n)} (\xi_F) &\sim &(1 - \xi_F^2)^{(c_F-n)/2} 
  {_2} F_1 \left(-n, 2c_F-n+1; c_F-n+1; \frac{1-\xi_F}{2} \right)\, ,
\nonumber \\
  n &=& 0,1,2,\cdots <c_F \, ,
\end{eqnarray}
where
\begin{eqnarray}
  \xi_F = \tanh [m_s (x^5 - l) ]\, ,\quad 
  c_F = \frac{\sqrt{2}\lambda_S v_S}{M_S} > 0\, .
\end{eqnarray}
The right-handed zero mode is projected out and their KK modes start
from level-one,
\begin{eqnarray}
  f_R^{(n)} (\xi_F) & \sim & (1 - \xi_F^2)^{(c_F-n)/2} 
  {_2} F_1 \left(-n+1, 2c_F-n; c_F-n+1; \frac{1-\xi_F}{2} \right)\, ,
  \nonumber \\
  n & = & 1,2,3, \cdots <c_F\, .
\end{eqnarray}
At each KK level except for the zero mode, fermions are vector-like
and the KK Dirac mass is given by
\begin{eqnarray}
  M_F^{(n)} = m_s \sqrt{2 n c_F - n^2}\, .
\end{eqnarray}
The profiles of the zero mode wave functions of the $X$-bosons, SM
gauge bosons and fermions are schematically shown in
Fig. \ref{fig:profile}.  The zero modes of the $X$-bosons (fermions)
are localized at the zero of $\langle \Sigma_Y \rangle$ ($\langle S
\rangle$) while the SM gauge bosons freely propagate in the bulk of
the extra dimension.

\begin{figure}[t!]
    \includegraphics[scale=0.7]{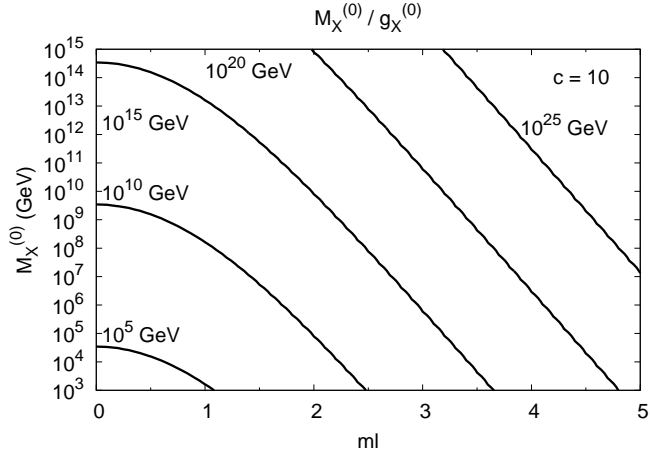}
    \caption{\label{fig:suppression} \footnotesize Contour of the
      effective $X$-boson mass scale $M_X^{(0)}/g_X^{(0)}$ in the
      $(ml, M_X^{(0)})$ plane.  We take
      $f_L^{(0)}(x^5)=f_X^{(0)}(x^5-l)$, $c=10$ and $g_5\sqrt{cm}=1$.
      Non-observation of proton decay requires $M_X^{(0)}/g_X^{(0)}
      \gtrsim 10^{15}~{\rm GeV}$\, .}
\end{figure}

Now we are at the stage to discuss the way to suppress the
four-dimensional couplings responsible for proton decay.  By
separating the zero modes of the $X$-bosons from those of the SM
fermions, the gauge coupling constants are exponentially suppressed
due to small overlaps of their wave functions:
\begin{eqnarray}
  g_X^{(0)} = g_5 \int_\infty^\infty \!\! {\rm d}x^5 \ 
  \left[ f_L^{(0)}(x^5) \right]^2 f_X^{(0)}(x^5)\, .
\end{eqnarray}
It is natural to assume that the five-dimensional gauge coupling
is related to the cutoff scale of this model as $g_5 \simeq
1/\sqrt{\Lambda}$.  Let us estimate the effective $X$-boson mass scale
${M_X^{(0)}}/{g_X^{(0)}}$ in the case where the cutoff scale is set to
the highest possible value: $\Lambda \simeq 5 g_5V \simeq cm$.  Even
if the $X$-boson mass itself is not large, ${M_X^{(0)}}/{g_X^{(0)}}$
can be significantly enhanced.  Figure \ref{fig:suppression} shows the
contour of $M_X^{(0)}/g_X^{(0)}$ in the $(ml, M_X^{(0)})$ plane.
Here, we assume the universal profile $f_L^{(0)}(x^5)=f_X^{(0)}(x^5 -
l)$, and take $c=10$ and $g_5\sqrt{cm}=1$.  For $ml \gtrsim 4$,
$X^{(0)}$-boson-mediated proton decay is sufficiently suppressed even
when $M_{X^{(0)}} \sim {\cal O}(1)$ TeV (See Eq.(\ref{eq:bound})).
Exchange of higher KK modes of the $X$-bosons $X^{(n)}$ can also
contribute to proton decay.  Such contributions can be suppressed to a
negligible level by a larger separation between the fermion zero modes
and $X^{(n)}$.

Let us discuss the case where the mass scale of the new particles of
our model is close to the current LHC reach.  Since the
five-dimensional SM gauge bosons have no mass term, the masses of
their first KK modes are given by the inverse of the size of the extra
dimension $L^{-1}$.  The fact that the LHC has already pushed the
limits on colored particle masses to ${\cal O}(1)$ TeV is translated
into the bound on the compactification scale $L^{-1} \sim 1$ TeV.
From Fig.\ref{fig:suppression}, the proton decay rate is adequately
suppressed if the distance between the SM fermions and the
$X^{(0)}$-bosons in the extra dimension satisfies the relation of $4
/m \lesssim l < L$.  These discussions require $M_X^{(0)} \gtrsim 10$
TeV and $M_F^{(1)} \gtrsim 10$ TeV for $c=10$.  Therefore, our model
predicts that the $X$-bosons or the first KK modes of the SM fermions
are not directly accessible at the LHC even when the first KK modes of
the SM gauge bosons are discovered.  The effects of the existence of
the terascale $X$-bosons may be indirectly observed through future
precision measurements at the International Linear Collider.  To
directly probe the structure of our model, colliders with higher
energies are necessary.

In GUTs, the electroweak Higgs doublet necessarily has a colored Higgs
triplet.  The SM fermions must couple to the Higgs doublet to acquire
mass while their couplings to the colored Higgs triplet should be
extremely small to avoid colored-Higgs-induced proton decay.  For
these reasons, different profiles between the Higgs doublet and
triplet are needed.  Such a difference can be achieved by using the
same mechanism as we have applied to the GUT gauge bosons.  In
addition to boson exchange diagrams, dimension six proton decay
operators suppressed by the inverse of the cutoff scale may exist.
These dimension six operators can be avoided by separating quarks and
leptons in the extra dimensions.  It is also worth investigating the
possibilities of generating the realistic fermion masses and mixing
angles using the wave function overlaps in the context of our model
\cite{Mirabelli:1999ks,KY2}.

The running of the gauge coupling constants is accelerated above the
compactification scale due to the existence of a tower of KK modes,
and the gauge coupling constants may be unified at a very low energy
scale \cite{Dienes}.  It is beyond the scope of this letter to
analyze the effects of KK modes on the gauge coupling unification.

In this letter, we have proposed a new extra-dimensional mechanism to
suppress the $X$-boson-induced proton decay rate to an experimentally
acceptable level.  The kink configuration of the vev of the adjoint
Higgs multiplet which breaks the grand unified symmetry allows
localization of the KK modes of the $X$-bosons.  The exponentially
small overlaps of the wave functions of the KK $X$-bosons and those of
the SM fermions are achieved by the separation of the $X$-boson and
fermion fields in the extra dimension, and stabilize the proton.
Although we have concentrated on the simplest model of an $SU(5)$ GUT
with one extra dimension, the same idea is applicable to other gauge
groups or extra dimension models.


\end{document}